\documentclass[3p,times]{elsarticle}

\usepackage{ecrc}

\volume{00}

\firstpage{1}

\journalname{Nuclear Physics A}

\runauth{A.D. Frawley on behalf of the PHENIX Collaboration}

\jid{nupha}

\usepackage{amssymb}
\usepackage{lineno}

\usepackage[figuresright]{rotating}

\begin{document}

\begin{frontmatter}



\dochead{}

\title{Cold Nuclear Matter Effects and Heavy Quark Production in PHENIX}


\author{A.D. Frawley on behalf of the PHENIX Collaboration}

\address{Physics Department, Florida State University, Tallahassee, FL 32306, USA}

\begin{abstract}
The PHENIX experiment uses semileptonic decay channels to measure open and closed heavy flavor cross sections across the rapidity range $-2.2 < y < 2.4$. 
High luminosity data are now available for p+p, d+Au, Cu+Cu and Au+Au collisions at $\sqrt{s_{NN}}$=200 GeV, and for Au+Au collisions at $\sqrt{s_{NN}}$= 62 GeV.
We discuss recent d+Au results for open heavy flavor and $J/\psi$ production, and discuss their implications for the cold nuclear matter contributions to 
heavy flavor production in heavy ion collisions.

\end{abstract}

\begin{keyword}


\end{keyword}

\end{frontmatter}

\linenumbers

\section{ Introduction}
\label{intro}

At low Bjorken momentum fraction ($x$) in nuclei, where gluon densities are very high, saturation effects are expected to become important~\cite{Gelis:2010nm}.
The modification of gluon densities in high energy collisions involving nuclear targets is inherently interesting because it contains information
about such gluon saturation effects, and also because it determines the initial conditions in high energy nucleus-nucleus collisions, where the hot matter
effects can be understood only after the initial conditions are known. 

Charm and bottom quarks are attractive as probes of the hot matter created in nucleus-nucleus collisions because their large mass prevents them from 
being created thermally in the hot medium. They are created only in hard processes that occur during the nuclear crossing, which is short in high energy 
collisions due to the large Lorentz contraction of the colliding nuclei in the collision frame. At RHIC, for example, the crossing time is $\sim 0.13 $  fm/$c$, which is 
smaller than the thermalization time of the QGP. 

The use of heavy quarks and quarkonia to probe the hot matter created in nuclear collisions would be straightforward if heavy flavor production scaled with the 
number of nucleon-nucleon collisions, because then the initial distributions could be obtained from measurements in $p+p$ collisions.
However there are a number of effects in $p$+A collisions that modify the initial distributions of open heavy flavor and quarkonia, including shadowing. These are
called cold nuclear matter (CNM) effects. They have to be studied using $p$+A data (or at RHIC, for technical reasons, $d$+A data). For this reason, PHENIX has
carried out a program of measurements that includes so far $p+p$, $d$+Au, Cu+Cu, Au+Au and, very recently, Cu+Au collisions.

\section{d+Au results}
\label{dAu}

\begin{figure}[htb!]
\centering
 \includegraphics[width=0.6\linewidth]{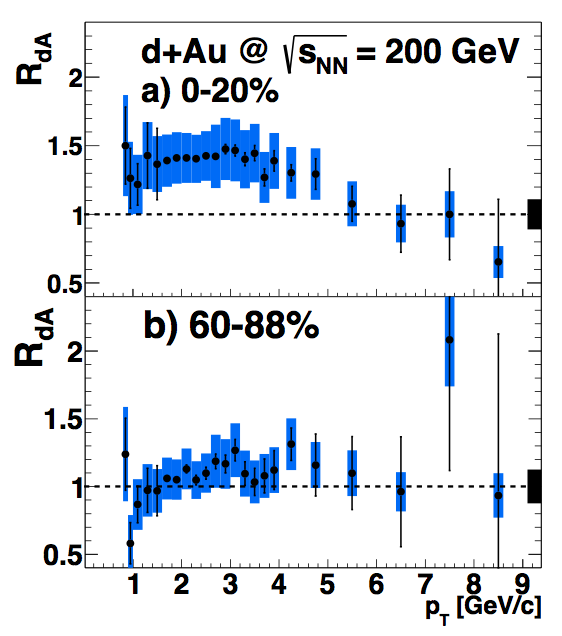}	
\caption{ The $R_{dAu}$ for HF electrons measured at $\sqrt{s_{NN}}= 200$ GeV for a) the most central collisions and for b) the most peripheral collisions~\cite{Adare:2012qb}.}
 \label{fig:hf_central}
\end{figure}

PHENIX has recently released new results on open heavy flavor (HF) production in $d$+Au collisions measured via semileptonic decays at mid rapidity~\cite{Adare:2012qb}. The $p_{\rm T}$ 
dependence of the nuclear modification is shown in Fig.~\ref{fig:hf_central} for central and peripheral collisions. Enhancement is observed for the 0-20\% most central collisions
in the $p_{\rm T}$ range 1-5 GeV/$c$, while little or no enhancement is observed for peripheral collisions. At high $p_{\rm T}$, where strong suppression is observed for 
HF electrons from Au+Au collisions, the measured $R_{dAu}$ is consistent with one. Thus the suppression observed in Au+Au collisions can be attributed, within
the uncertainties on $R_{dAu}$, to hot nuclear matter effects.

\begin{figure}[htb!]
\centering
 \includegraphics[width=0.6\linewidth]{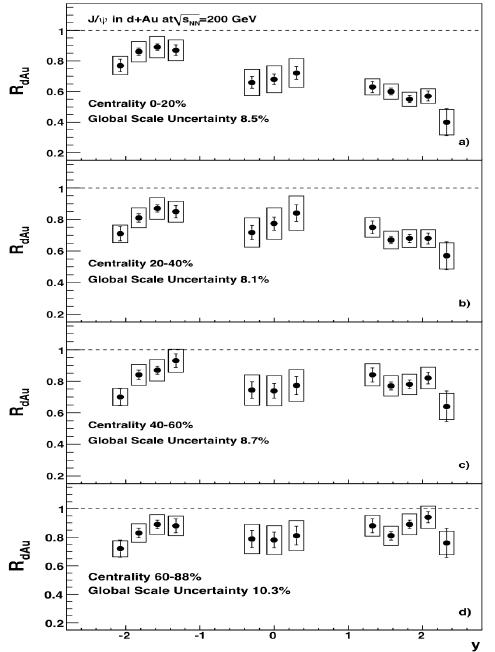}	
\caption{ $J/\psi$ $R_{dAu}$ versus rapidity for four centrality bins~\cite{Adare:2010fn}. }
 \label{fig:jpsi_rap_4cent}
\end{figure}

\begin{figure}[htb!]
\centering
 \includegraphics[width=0.6\linewidth]{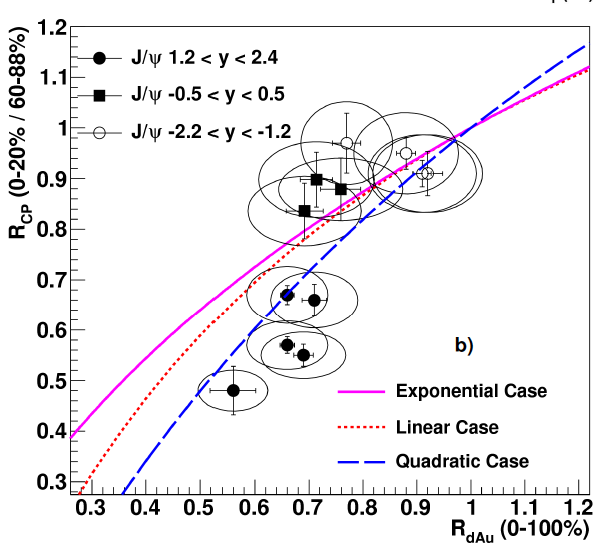}	
\caption{$R_{dAu}$ plotted versus $R_{CP}$~\cite{Adare:2010fn}. The data points show the location for the 12 rapidity bins measured by PHENIX, the ellipses represent 
the systematic uncertainties. See the text for discussion. }
 \label{fig:rcp_vs_rdau_jpsi}
\end{figure}

Published PHENIX $d$+Au $J/\psi$ data from the 2008 RHIC run span the rapidity range -2.2 to 2.4 in 12 rapidity bins, and have been divided into
4 centrality bins~\cite{Adare:2010fn}. Fig.~\ref{fig:jpsi_rap_4cent} shows the $J/\psi$ $R_{dAu}$ as a function of rapidity for each of the four centrality bins.
The forward rapidity data for the most central collisions show suppression of about 50\%.  The centrality dependence of the forward rapidity data reveals
an interesting feature, which is illustrated in Fig.~\ref{fig:rcp_vs_rdau_jpsi}~\cite{Adare:2010fn}. In this plot the centrality integrated $R_{dAu}$ on the horizontal axis represents the 
overall modification. The ratio of the modification for the most central and most peripheral collisions, $R_{CP}$, on the vertical axis characterizes the centrality slope of the 
modification. The points show the locations of the PHENIX
measurements at twelve rapidities, with the systematic uncertainties represented by ellipses. The lines on the plot are loci generated from a Glauber model of the
$d$+Au collisions in which the modification for each nucleon-nucleus collision is assumed to have certain simple mathematical dependencies on the nuclear 
thickness $T_A(r_T)$ at the nucleon impact parameter $r_T$. The three mathematical forms shown are exponential, linear and quadratic. The plot shows that the 
forward rapidity modifications require a dependence on $T_A(r_T)$ that is quadratic, or greater.

\begin{figure}[htb]
\centering
  \includegraphics[width=0.6\linewidth]{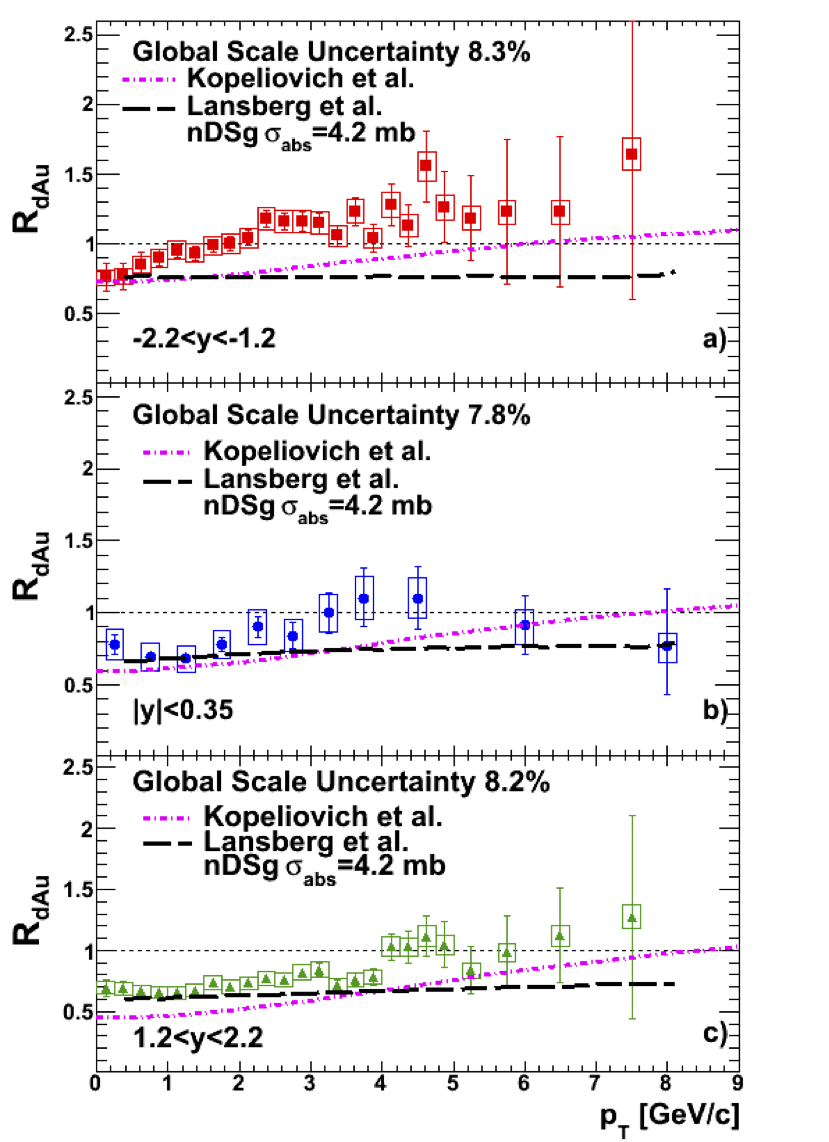}
 \caption{ The  $J/\psi$ $R_{dAu}$ versus $p_{\rm T}$ at backward, mid and forward rapidity~\cite{Adare:2012qf}. The theory curves are described in the text.}
 \label{fig:rdau_mb_pt}
\end{figure}

PHENIX has recently released measurements of the $p_{\rm T}$ dependence of the $J/\psi$ $R_{dAu}$ as a function of centrality~\cite{Adare:2012qf}. 
The data are compared in~\cite{Adare:2012qf} with various model calculations. A comparison of the centrality integrated data with two 
models is shown here in Fig.~\ref{fig:rdau_mb_pt}. One calculation~\cite{Kopeliovich:2011zz}, shown as the dot-dashed line, uses a  model of the color 
dipole breakup based on a parameterization of HERA data, a Cronin broadening that is parameterized from low energy data, and a shadowing correction 
obtained from the nDSG nuclear parton distribution function (nPDF) parameterization. The calculation over-predicts the suppression, but otherwise reproduces the 
shape of the $p_{\rm T}$ distribution at forward and mid rapidity. However it does a poor job of reproducing the shape of the distribution at backward rapidity.
In the second calculation~\cite{Ferreiro:2012zb,Ferreiro:2008wc}  shadowing is parameterized using various nPDF sets, and the effect of collisions of the $c\bar{c}$ 
dipole with nucleons during the nuclear crossing is represented by an effective breakup cross section. The example shown in Fig.~\ref{fig:rdau_mb_pt} as the dashed line 
uses the nDSG nPDF set, with an effective breakup cross section of 4.2 mb. There is no added Cronin enhancement. This second calculation describes the data 
reasonably well at low $p_{\rm T}$ at forward and mid rapidity, although it under-predicts the data at higher $p_{\rm T}$. However at backward rapidity this calculation 
also fails.

At backward rapidity and low $p_{\rm T}$ (where $x \sim 0.1$ for the parton in the Au nucleus) production occurs in the anti-shadowing region, while at high $p_{\rm T}$ ($x \sim 0.3$) 
production moves towards the EMC region. No direct evidence of an EMC effect has been reported in the gluon distributions, and few constraints on the gluon distributions 
exist in this region. Furthermore there is large disagreement in the modification of the gluon density between nPDFÕs. The nDSg nPDF set includes no suppression in the EMC region, and 
only a small anti-shadowing effect, while the EKS98 nPDF exhibits a suppression similar to that observed in the quark distributions, and a larger anti-shadowing effect. 
The lack of a strong anti-shadowing effect combined with the absence of an EMC effect in the nDSg nPDF causes the calculated $R_{dAu}$ to remain roughly constant with increasing 
$p_{\rm T}$. For the EKS98 nPDF set, the larger anti-shadowing combined with the inclusion of an EMC effect causes a decrease in the calculated $R_{dAu}$  as $p_{\rm T}$ 
(and correspondingly $x$) increases~\cite{Adare:2012qf}, worsening agreement with the data. The explanation of the $p_{\rm T}$ shape at backward rapidity remains unclear. 
The inclusion of a Cronin effect in~\cite{Kopeliovich:2011zz} that seems adequate to describe the more forward rapidity data does not allow a good description of the backward 
rapidity data, suggesting that the disagreement may be due to the nPDF sets at high $x$, or possibly to some other physics effect not considered.

\begin{figure}[htb!]
\centering
 \includegraphics[width=0.7\linewidth]{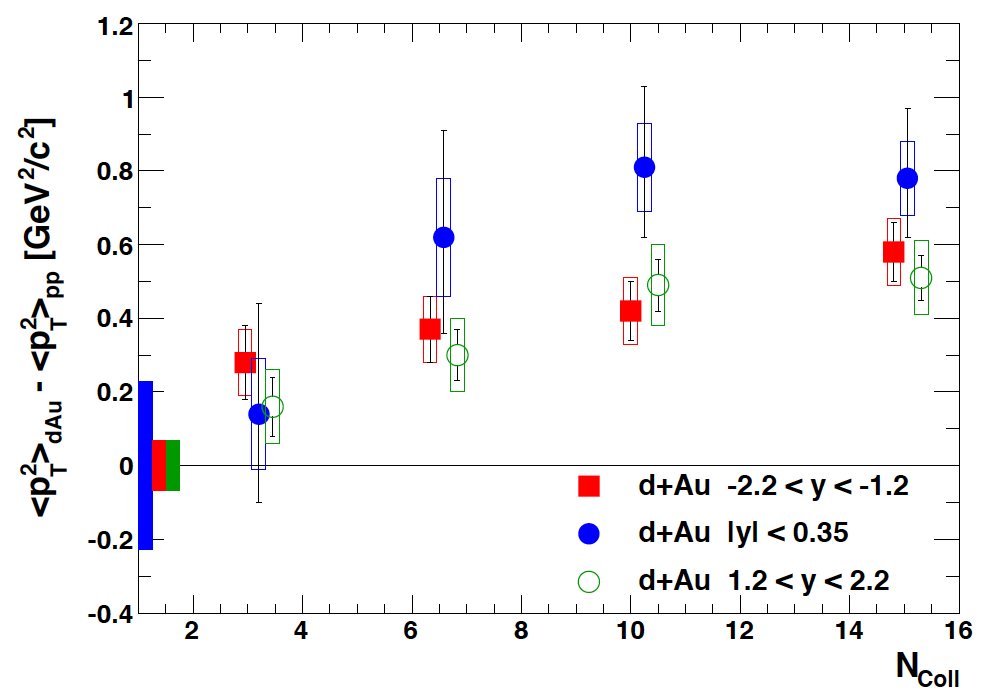}	
\caption{The enhancement of $\langle p^{2}_{\rm T} \rangle$ for $d$+Au collisions over $p+p$ collisions, plotted versus collision centrality at backward, mid and forward rapidity~\cite{Adare:2012qf}. }
 \label{fig:mean_pt_sq}
\end{figure}

The $J/\psi$ $p_{\rm T}$ broadening relative to $p+p$ collisions can be characterized by plotting the difference of the values of $\langle p^{2}_{\rm T} \rangle$ between the $d$+Au
and $p+p$ distributions. This difference is shown as a function of centrality, characterized by $\langle N_{coll} \rangle$, in Fig.~\ref{fig:mean_pt_sq}. The 
dependence on centrality is very similar at forward and backward rapidity, with a hint that it is larger at mid rapidity. However it should be noted that $\langle p^{2}_{\rm T} \rangle$
is about 20\% larger at mid rapidity than it is at forward or backward rapidity for both $p+p$ and $d$+Au collisions~\cite{Adare:2012qf}, so that the difference plotted in 
Fig.~\ref{fig:mean_pt_sq} roughly scales with the size of $\langle p^{2}_{\rm T} \rangle$.

%$\Upsilon$ $R_{dAu}$ rapidity distribution, with $J/\psi$ comparison (with STAR midrapidity point?).
%\begin{figure}[htb!]
% \includegraphics[width=0.8\linewidth]{./figures/rdau_vs_rapidity_upsilons_jpsi}	
%\caption{  }
% \label{fig:upsilon_dau_rap}
%\end{figure}

\section{Heavy Ion Results}
\label{HI}

There are new semileptonic decay open heavy flavor PHENIX results for heavy ions that include Cu+Cu $R_{AA}$ data at both mid and forward rapidity. 

\begin{figure}[htbl]
\begin{minipage}[b]{0.45\linewidth}
\centering
\includegraphics[width=\textwidth,height=\textwidth]{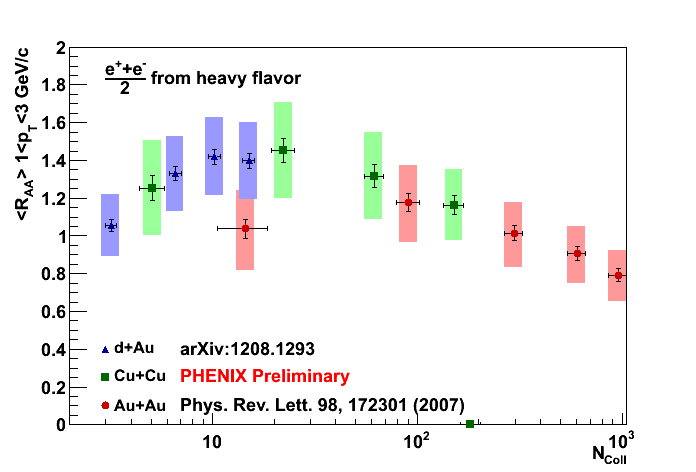}
\caption{Heavy flavor electron modifications at mid rapidity for Cu+Cu collisions compared with those for $d$+Au and Au+Au. In all cases the modifications are integrated over
the range $p_{\rm T}=1-3$ GeV/$c$.}
\label{fig:hf_central_lo}
\end{minipage}
\hspace{0.5cm}
\begin{minipage}[b]{0.45\linewidth}
\centering
\includegraphics[width=\textwidth,height=\textwidth]{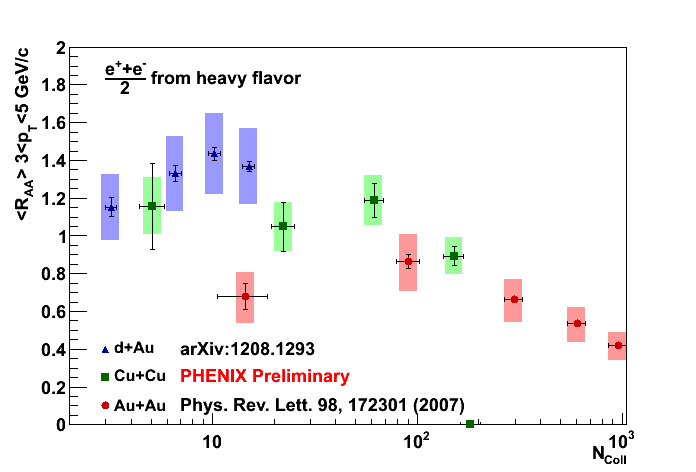}
\caption{Heavy flavor electron modifications at mid rapidity for Cu+Cu collisions compared with those for $d$+Au and Au+Au. In all cases the modifications are integrated over
the range $p_{\rm T}=3-5$ GeV/$c$.}
\label{fig:hf_central_hi}
\end{minipage}
\end{figure}

In Fig.~\ref{fig:hf_central_lo} and Fig.~\ref{fig:hf_central_hi} the mid rapidity Cu+Cu $R_{AA}$ data are compared as a function of $N_{coll}$ with 
data from $d$+Au and Au+Au collisions. The data are shown for two $p_{\rm T}$ ranges. Where they overlap, the three data sets are found to display similar behavior 
with $N_{coll}$ within uncertainties. It is noteworthy that the the $d$+Au data and the peripheral Cu+Cu data have very similar modifications, including some 
enhancement at $N_{coll} \sim 10$.

\begin{figure}[htb!]
\centering
 \includegraphics[width=0.8\linewidth]{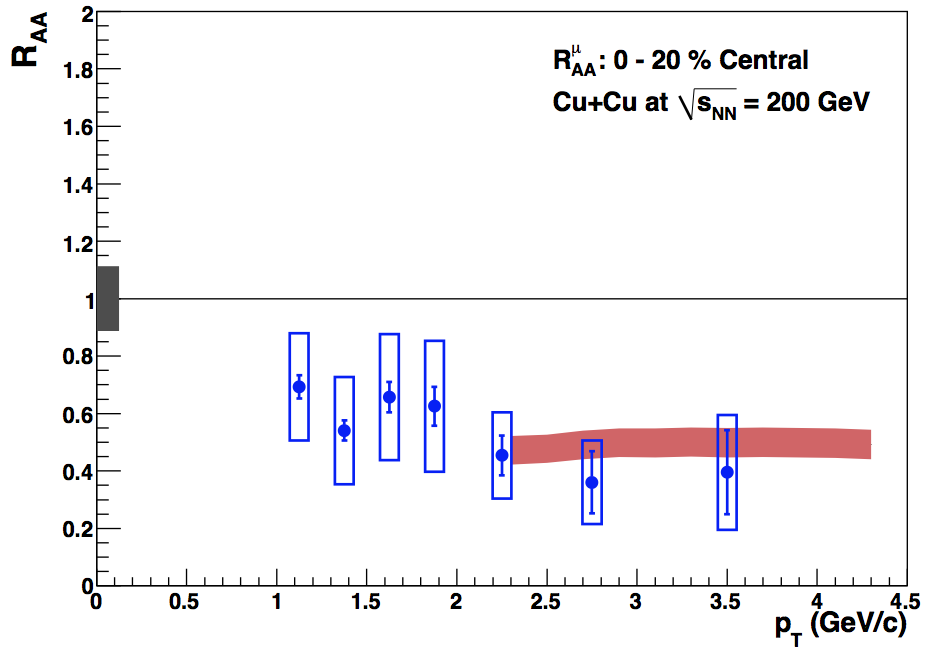} 
 \caption{ The $p_{\rm T}$ dependence of the $R_{AA}$ for semileptonic decays of heavy quarks to muons in Cu+Cu collisions at $1.4<|y|<1.9$~\cite{Adare:2012px}. The theory 
 curve~\cite{Sharma:2009hn} is discussed in the text.}
 \label{fig:cucu_hf_muons}
\end{figure}

At forward rapidity ($1.4<|y|<1.9$) , recent results on the $R_{AA}$ for semileptonic decays of heavy quarks to muons in Cu+Cu collisions have now been released~\cite{Adare:2012px}. The 
$R_{AA}$ for the most central collisions is shown in Fig.~\ref{fig:cucu_hf_muons}. The suppression in central collisions is found to be stronger than that seen at comparable 
$N_{part}$ at mid rapidity for Au+Au collisions, suggesting the possibility that CNM effects are larger at forward rapidity. The observed suppression is consistent with 
a calculation~\cite{Sharma:2009hn} that includes the effects of heavy-quark energy loss 
and in-medium heavy meson dissociation, as well as cold nuclear matter effects due to shadowing and initial state energy loss due to multiple scattering of incoming partons before 
they interact to form the $c\bar{c}$ pair.

%$J/\psi$ global plot of survival vs $dN/d\eta$. 
%\begin{figure}[htb!]
%\centering
% \includegraphics[width=0.6\linewidth]{./figures/PHENIX_NA60_survival_multiplicity}	
%\caption{ }
% \label{fig:jpsi_survival_global}
%\end{figure}

\begin{figure}[htb!]
\centering
\includegraphics[width=0.6\linewidth]{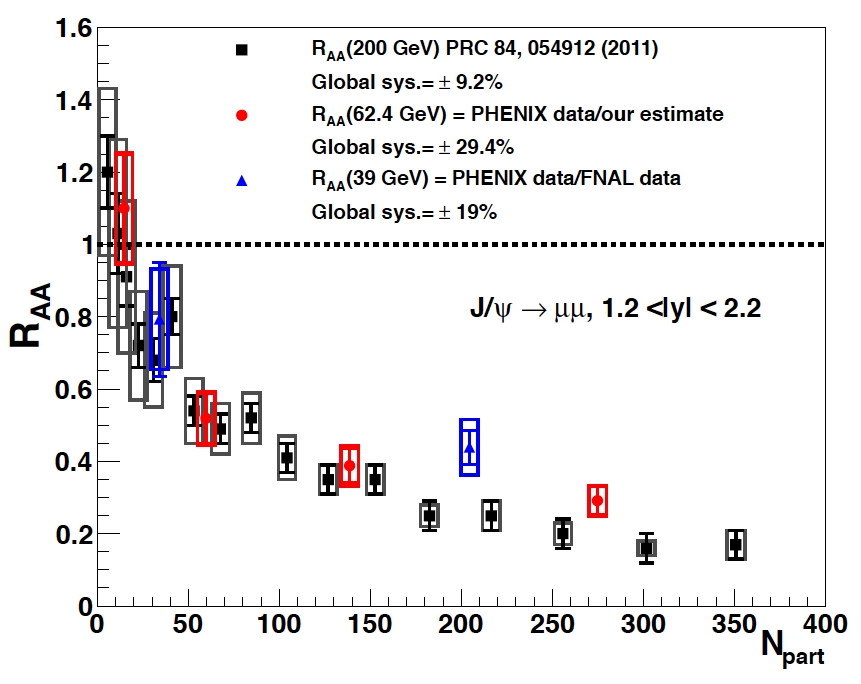}	
\caption{Comparison of newly released Au+Au $J/\psi$ $R_{AA}$ data at 62 and 39 GeV with existing data at 200 GeV.}
 \label{fig:jpsi_raa_3energies}
\end{figure}

PHENIX has recently released measurements of the $J/\psi$ $R_{AA}$ at $\sqrt{s_{NN}}$~=~62 and 39 GeV in the rapidity range $1.2<|y|<2.2$~\cite{Adare:2012wf}. These are 
compared in Fig.~\ref{fig:jpsi_raa_3energies} with the existing 200 GeV data~\cite{Adare:2011yf}. The modifications are similar at the three energies, but it should be noted  
that CNM effects are expected to be quite different for $J/\psi$ production at the three energies~\cite{Lourenco:2008sk}. We do not yet have comparison $d$+Au data at 
62 and 39 GeV. 

Taken together, the ALICE Pb+Pb data at $\sqrt{s_{NN}}=2.76$ TeV~\cite{Abelev:2012rv}, the NA50 Pb+Pb data at $\sqrt{s_{NN}}=17.3$ GeV~\cite{Arnaldi:2009ph}, and the PHENIX 
Au+Au data at 39, 62 and 200 GeV provide a $J/\psi$ data set that spans a large energy density range and a broad range of initial conditions and charm production cross sections. 
At sufficiently high $c\bar{c}$ pair production rates, a significant fraction of the $J/\psi$ yield is expected to be due to coalescence of $c$ and $\bar{c}$ quarks from different hard 
processes in the same nuclear collision, and comparison of data at LHC and RHIC energies will be necessary to constrain this contribution.  
Once the necessary CNM reference data are
obtained for the PHENIX 39 and 62 GeV data, and the ALICE 2.76 TeV data, the available data should provide strong constraints on theories of $J/\psi$ modification in heavy ion collisions.

\section{Summary}

Recently released PHENIX $R_{dAu}$ data at 200 GeV for HF electrons at mid rapidity show enhancement for central collisions in the range $p_{\rm T}=1-5$ GeV/$c$, with no 
evidence of significant modification at higher $p_{\rm T}$. These results indicate that the strong HF electron suppression observed for Au+Au collisions at medium to 
high $p_{\rm T}$ is due to hot matter effects.

New results have now been released for the centrality and $p_{\rm T}$ dependence of $R_{dAu}$ for $J/\psi$ production at 200 GeV. The data were measured in rapidity intervals 
$-2.2 < y < -1.2$,  $|y| < 0.35$ and $1.2 < y < 2.2$. Together with previously published $R_{dAu}$ vs centrality data at twelve rapidities, these complete the PHENIX $J/\psi$ $R_{dAu}$ 
results from the 2008 RHIC run. Two notable observations have been made about the data. First, the centrality dependence of the modification at the most forward rapidities is found 
to be stronger than linear in the nuclear thickness. Second, the $p_{\rm T}$ dependence of $R_{dAu}$ at backward rapidity is not described by available calculations, suggesting 
perhaps a problem with the nPDF parameterizations for $x$ values in the anti-shadowing and EMC region. 

There are new PHENIX preliminary HF electron $R_{AA}$ data from 200 GeV Cu+Cu collisions. Where they overlap, $d$+Au, Cu+Cu and Au+Au show very similar behavior with $N_{coll}$,
including enhancement of the $d$+Au and Cu+Cu data at $N_{coll} \sim 10$, at lower $p_{\rm T}$. Newly released HF muon $R_{AA}$ data from 200 GeV Cu+Cu collisions show 
suppression that is stronger than is seen at mid rapidity at the same $N_{part}$, perhaps consistent with stronger CNM effects at forward rapidity.

PHENIX has now released $J/\psi$ $R_{AA}$ data from 39 and 62 GeV Au+Au collisions. The modifications are found to be similar to that at 200 GeV. However CNM effects are 
known to be strongly energy dependent, and CNM reference data will be needed before conclusions can be drawn about the differences between hot matter effects at these energies.



%% References


\bibliographystyle{elsarticle-num}
\bibliography{frawley_hp2012}






% \end{thebibliography}

\end{document}